\documentclass[12pt]{article}
\usepackage{amsmath}
\usepackage{latexsym}
\usepackage{graphicx}
\usepackage{kotex}
\DeclareGraphicsExtensions{.jpg,.pdf,.png,.gif}
\usepackage{amssymb}
\newtheorem{remark}{Remark}
\usepackage{multirow}

\def\bfX{\mbox{\boldmath{$X$}}}

\def\bfbeta{\mbox{\boldmath{$\beta$}}}
\def\bflam{\mbox{\boldmath{$\lambda$}}}

\def\no{\noindent}

\newcommand{\bx}{\mbox{\boldmath{$x$}}}

\newcommand{\Ti}{\mathrm{\scriptscriptstyle T}}

\begin{document}

\no {\Large {\bf Sampling techniques for big data analysis in finite population inference}}

\bigskip

\bigskip

\no
{\large {\bf  Jae Kwang  Kim and Zhonglei Wang}}

\bigskip

{\em
\no
Department of Statistics, Iowa State University, Ames, Iowa 50011, U.S.A.

\no
E-mails: jkim@iastate.edu \ \ and \ \ wangzl@iastate.edu
}

\bigskip

\bigskip

\no
{\bf Summary}

\medskip

{\bf In analyzing big data for finite population inference,  it is critical to adjust for the
	selection bias in the big data. In this paper, we propose two methods of reducing the selection bias
	associated with the big data sample. The first method uses a version of inverse sampling by incorporating auxiliary information from external sources, and the second one borrows the idea of data integration by combining the big data sample with an independent probability sample.  Two simulation studies show that the proposed methods are unbiased and have better coverage rates than their alternatives. In addition, the proposed methods are easy to implement  in practice.
}

\medskip

\no

\bigskip

\bigskip

\section{Introduction}

Probability sampling is a scientific tool for obtaining a representative sample from a target finite population. Formally, a probability sample has the property that every element in the finite population has a known and nonzero probability of being selected. Probability sampling can be used to construct valid statistical inferences for finite population parameters.  Survey sampling is an area of statistics that deals with constructing efficient probability sampling designs and corresponding estimators. Classical approaches in survey sampling are discussed in Cochran (1977),  S\"{a}rndal {\em et al.} (1992)  and Fuller (2009).

Despite the merits of probability samples, Baker {\em et al.} (2013) argue that it becomes common to get non-probability samples, which may not represent the target population properly. Besides, collecting  a strict probability sample is almost impossible in certain areas due to unavoidable issues such as frame undercoverage and nonresponse. The increasing prevalence of non-probability samples, such as web panels, makes methods for non-probability samples even more important.
Keiding and Louis
(2016) address  the challenges in using non-probability samples for making inferences.
Elliott and Valliant (2017) review the weighting methods for reducing the selection bias in non-probability samples.
Rivers (2007) proposes  nearest neighbor imputation matching for combining information from survey data and big data.
Bethlehem
(2016) discusses sample matching methods for handling non-probability samples.

%Keiding, N. and Louis, T. A. (2016). Perils and potentials of self-selected entry to epidemiological studies and surveys (with Discussions). Journal of the Royal Statistical Society. Series A. 179 1-28.

Big data is one example of such non-probability sample. The Four Vs (volume, velocity, variety and veracity) of big data and its implication to statistical inference are nicely discussed in Franke {\em et al.} (2016).
While use of big data for predictive analysis is a hot area of research (Efron and Hastie, 2016), its use for  finite population inference is not well investigated in the literature.  Tam (2015) discusses a statistical framework for analyzing big data for official statistics,  particularly in agricultural statistics.   Rao and Molina (2015) discuss using the area-level summary of big data as one of the covariates in the linking model for small area estimation.
Tam and Kim
(2018) cover some ethical challenges of big data for official statisticians and discuss some preliminary methods of correcting for selection bias in  big data.

One of the benefits of using big data  is, as pointed out by  Tam and Clarke (2015),
% Tam and Clark (2015),
in the cost effectiveness in
the production of official statistics.
However, there are still great challenges when using big data for finite population inference.
The most critical issue is  how to handle selection bias in the big data sample (Meng, 2018). Adjusting for  the selection bias in big data is an important practical problem in survey sampling.

In this paper, we discuss how  some of the sampling techniques can be applied in harnessing    big data for finite population inference.
By treating the selection bias in the big data sample as a missing data problem, we propose two approach of handling big data in survey sampling. The first approach is based on inverse sampling, which is
a special case of two-phase sampling,  and a novel inverse sampling method is proposed  to obtain a representative sample from the big data. The second approach  is based on the weighting method using the  auxiliary information obtained from another independent probability sample. Combining information from two data sources, often called data integration, is also a hot area of research in survey sampling.  In the proposed method, an independent probability sample is used to estimate the parameters of the propensity score model for the big data sample.

The paper is organized as follows. In Section 2, the basic setup is introduced and the selection bias of big data is discussed. In Section 3, an inverse sampling method is proposed. In Section 4, a propensity score weighting approach using data integration is discussed. Results from two limited simulation studies are presented in Section 5. Some concluding remarks are made in Section 6.

%[Some description of big data can be placed here. ]

\section{Basic Setup}

Consider a finite population $\{y_i:i\in U\}$, where $y_i$ is the $i$-th observation of the study variable $Y$, and $U=\{1,\ldots,N\}$ is the corresponding index set with known size $N$. A big data sample $\{y_i:i\in B\}$ is available with $B\subset U$. Specifically, $\delta_i = 1$ if $i \in B$ and $\delta_i=0$ otherwise, and assume that  $y_i$ is observed only when $\delta_i=1$.
We are interested in estimating the population mean $\bar{Y}_N= N^{-1} \sum_{i=1}^N y_i$.

From  the big data sample $B$, we can estimate $\bar{Y}_N$ by $\bar{Y}_B = N_B^{-1} \sum_{ i=1}^N \delta_i y_i$, where $N_B = \sum_{i=1}^N \delta_i$ is the known size of $B$.  Given $\{\delta_i:i\in U\}$, the error of $\bar{Y}_B$  can be written as
\begin{equation*}
\bar{Y}_B - \bar{Y}_N = \frac{1}{f_B} \mbox{Cov}( \delta, Y)
\label{1}
\end{equation*}
where $f_B=N_B/N$ and
$$ \mbox{Cov} (\delta, Y) = \frac{1}{N} \sum_{i=1}^N ( \delta_i - \bar{\delta}_N ) ( y_i - \bar{Y}_N)
$$
with $\bar{\delta}_N = N^{-1} \sum_{i=1}^N \delta_i $. Thus, we have
\begin{equation}
E_\delta\{ ( \bar{Y}_B - \bar{Y}_N )^2 \}  = \frac{1}{f_B^2 }E_\delta \left\{ \mbox{Cov} (\delta, Y)^2 \right\}\,, \label{2}
\end{equation}
where $E_{\delta} ( \cdot)$ denotes the expectation with respect to the random mechanism for $\delta_i$.

If the random mechanism for $\delta_i$ is based on Bernoulli sampling, where the inclusion indicators follow a Bernoulli distribution with success probability $f_B$ independently, we can obtain
\begin{eqnarray*}
	E_\delta \left\{ \mbox{Cov} (\delta, Y)^2 \right\} &=& \left[ E_\delta \left\{ \mbox{Cov} (\delta, Y) \right\} \right]^2 +   \mbox{Var}_{\delta}\{  \mbox{Cov} (\delta, Y)  \} \\
	&=& 0 + {\frac{1}{N^2} \sum_{i=1}^N (y_i - \bar{Y}_N)^2 f_B (1-f_B) = \frac{1}{N} f_B (1- f_B) \sigma^2 }
\end{eqnarray*}
with $\sigma^2 = N^{-1} \sum_{i=1}^N ( y_i - \bar{Y}_N)^2 $. Thus, under Bernoulli sampling, (\ref{2}) reduces to
$$ E_\delta\{ ( \bar{Y}_B - \bar{Y}_N )^2 \} = \frac{1}{N_B} (1-f_B) \sigma^2\, , $$
which is consistent with the classical theory for Bernoulli sampling with sample size $n=N_B$. For general cases, (\ref{2}) can be expressed as
\begin{eqnarray}
E_\delta\{ ( \bar{Y}_B - \bar{Y}_N )^2 \}  &=&
\frac{1}{f_B^2 }E_\delta \left\{ \mbox{Corr} (\delta, Y)^2 \mbox{Var}( \delta)  \mbox{Var}( Y) \right\}  \notag \\
&=&  E_\delta \left\{ \mbox{Corr} (\delta, Y)^2  \right\} \times \left( \frac{1}{f_B} - 1 \right) \times  \sigma^2\,,  \label{3}
\end{eqnarray}
%\section{Reservior sampling}
where the second equality follows from
$$ \mbox{Var}( \delta) = \frac{1}{N} \sum_{i=1}^N ( \delta_i - \bar{\delta}_N)^2 = f_B ( 1-f_B)\,. $$

Equality (\ref{3}) is also presented in Meng (2018). Although there are three terms in (\ref{3}) determining the selection bias of $\bar{Y}_B$, the first term, $E_\delta \left\{ \mbox{Corr} (\delta, Y)^2  \right\}$, is the most critical one. Meng (2018) calls the term \emph{Data Defect Index} (DDI), which determines the level of departure from simple random sampling. Under equal probability sampling designs such that $E_\delta(\delta_i)=f_B$, we have   $E_\delta \{ \mbox{Corr} (\delta, Y) \}=0$ and DDI is of order $O(1/N)$, which implies  $E_{\delta}\{ ( \bar{Y}_B- \bar{Y}_N )^2 \}  = O(N_B^{-1})$. For other sampling designs with $E_\delta \{ \mbox{Corr} (\delta, Y) \} \neq  0$, the DDI becomes significant with order $O(1)$, which implies $E_{\delta}\{ ( \bar{Y}_B- \bar{Y}_N )^2 \} = O( N_B^{-1} N -1)$. Therefore, a non-probability sampling design with  $E_\delta \{ \mbox{Corr} (\delta, Y) \} \neq 0$ makes the analysis results subject to selection bias.

In this paper, we show how to use  some of the existing sampling techniques to reduce the selection bias of the big data sample and make the resulting analysis valid. We consider two techniques, one is  inverse sampling and the other is  survey data integration.
%Before we introduce two techniques, we first introduce the setup and the issue of selection bias in big data.

\section{Inverse sampling}

%Sampling techniques have been developed in statistics as a way of obtaining representative samples.
%\subsection{Introduction}
When the distribution of the study variable for the big data sample differs systematically from that for the target population, the big data sample does not necessarily represent the target population.  An important question in this respect is whether we can use  auxiliary variables, external to the big data sample, to correct for the selection bias.  In this section, we cosider a novel inverse sampling approach to address this problem. The proposed inverse sampling can be viewed as a special case of  two-phase sampling (e.g., Breidt and Fuller, 1993; Rao and Sitter, 1995; Hidiroglou,
2001; Kim, {\em et al.} 2006; Stukel and Kott, 1996). The first-phase sample is  the big data sample, which is subject to  selection bias. The second-phase sample is a subsample of the first-phase sample to correct  the selection bias of the big data sample. Inverse sampling is originally proposed as a way of obtaining a simple random sample from a sample obtained from a complex sampling design.  For some classical designs, such as stratified sampling, the inverse sampling algorithm is presented by Hinkins {\em et al.}
(1997) and Rao {\em et al.} (2003). Till\'{e} (2016) applies the inverse sampling concept to a quota sample.  We address the application of inverse sampling to  big data  subject to   selection bias.

Unlike the classical two-phase sampling, the first-phase sample in our setup is the big data itself, and we have no control over it. Thus, we first use some external source to determine the level of selection bias in the big data. This step can be called weighting step, as the importance weights are computed for each element in the big data sample. The second step is to select the second phase sample from the big data with the selection probability proportional to  the importance weights.

To correct for selection bias using the proposed inverse sampling approach, we need external information about the target population, either from a census or from a probability sample, for some auxiliary variable $\bx$. To formally present the idea, let $(\bx_i, y_i)$ be available in the big data sample ($B$) and  $f( \bx)$ be the density for the marginal distribution of $\bx$ that is obtained from an external source. We assume that the auxiliary variable $\bx$ has a finite second moment. We are interested in estimating $\theta=E(Y)$ from the big data sample $B$. The first-order inclusion probability for the big data sample $B$ is unknown.

Using the idea of importance sampling (Goffinet and Wallach, 1996; Henmi {\em et al.}
2007), it  can be shown that
\begin{equation}
\hat{\theta}_{B1}  = \frac{ \sum_{ i \in B} \frac{ f( \boldsymbol{x}_i)}{ f ( \boldsymbol{x}_i \mid \delta_i=1) } \frac{ f( y_i \mid\boldsymbol{x}_i) }{ f( y_i \mid \boldsymbol{x}_i , \delta_i=1) } y_i }{ \sum_{ i \in B}  \frac{ f( \boldsymbol{x}_i) }{ f( \boldsymbol{x}_i \mid \delta_i =1 ) }  \frac{ f( y_i \mid \boldsymbol{x}_i) }{ f( y_i \mid \boldsymbol{x}_i , \delta_i=1) } }\,,
\label{4}
\end{equation}
is asymptotically  unbiased for $\theta=E(Y)$ by assuming that $f(\delta_i=1\mid \bx_{i})>0$ for $i\in U$ almost surely.  If the sampling mechanism for $B$ is ignorable after controlling  on $\bx$, i.e. $P( \delta_i = 1 \mid \bx_i, y_i )=P( \delta_i =1 \mid \bx_i),$
then (\ref{4}) reduces to
\begin{equation}
\hat{\theta}_{B1} = \frac{\sum_{ i \in B}  \frac{ f( \boldsymbol{x}_i)}{ f(\boldsymbol{x}_i \mid \delta_i =1 ) }
	y_i }{ \sum_{ i \in B} \frac{ f( \boldsymbol{x}_i)}{ f( \boldsymbol{x}_i \mid \delta_i =1 ) }    }  := \sum_{i \in B} w_{i1} y_i\,.
\label{5}
\end{equation}
The weight $w_{i1}$ can be called importance weight, following the idea of importance sampling.
If $\bx_i$ is a vector of stratum indicator variables, then $f( \bx_i )/f( \bx_i \mid \delta_i =1)$   equals to $(N_h/N)/(n_h/n) $ for $i$ in stratum $h$, which leads to unbiased estimation under stratified sampling.

If  only $\bar{\bfX}_N=N^{-1}\sum_{i=1}^N\bx_i $ is available,  we can approximate $f(\bx)$ by $f_0(\bx)$, which
minimizes the  Kullback-Leibler   distance
\begin{equation}
\mbox{min}_{f_0 \in P_0} \int f_0 \left( \bx\right)\ln \left\{  \frac{f_0 \left( \bx \right)}{f \left(\bx \mid \delta = 1 \right) } \right\} \mbox{d} \bx\,,
\label{6}
\end{equation}
where  $P_0 = \{ f(\bx) ; \int \bx f(\bx) \mbox{d}\bx = \bar{\bfX}_N \}$.   The solution to (\ref{6}) is
\begin{equation}
f_0\left( \bx \right) = f\left( \bx \mid \delta=1 \right) \frac{ \mbox{exp} \left( \bx^\Ti{\bflam} \right)  }{E\left\{ \mbox{exp} \left( \bfX^\Ti{\bflam}\right) \mid \delta=1 \right\} }\,,
\label{7}
\end{equation}
where $\bflam$ satisfies $\int \bx f_0 \left(\bx \right) \mbox{d} \bx = \bar{\bfX}_N $, and $D^\Ti$ is the transpose of  $D$. Thus, the selection probability for the second-phase selection is proportional to $ \exp \left( \bx^\Ti{\bflam}\right)$, which is very close to the  exponential tilting   calibration discussed in Kim (2010).   Using (\ref{7}), the weighted estimator in (\ref{5}) reduces to
\begin{equation}
\hat{\theta}_{B1} = \frac{\sum_{ i \in B}   \exp (  \bx_i^\Ti\hat{\bflam})  y_i }{ \sum_{ i \in B}  \exp (\bx_i^\Ti\hat{\bflam})    }\,,
\label{8}
\end{equation}
where $\hat{\bflam}$ satisfies
\begin{equation}
\frac{\sum_{ i \in B}   \exp ( \bx_i^\Ti\hat{\bflam})  \bx_i }{ \sum_{ i \in B}  \exp (\bx_i^\Ti\hat{\bflam})    } = \bar{\bfX}_N\,.
\label{9}
\end{equation}
Here, equation (\ref{9}) can be called calibration equation (Wu and Sitter, 2001). Unlike the usual calibration estimation, we may ignore the sampling variability in estimating $\bflam$ since   $N_B$ is large. When the sample size of $B$ is large, the computation for calibration equation (\ref{9}) may be challenging. In this case, one-step approximation (Kim, 2010) can be used.

Based on (\ref{8}), we discuss how to select the second phase sample $(B_2)$ of size $n$ from the big data sample $B$ such that
$\hat{\theta}_{B2} =n^{-1} \sum_{i \in B_2} y_i$ is approximately design unbiased for $\hat{\theta}_{B1}$ in (\ref{5}). The basic idea is to choose  the conditional first-order inclusion probability $\pi_{i 2 \mid 1} = P ( i \in B_2 \mid i \in B)$  such that
\begin{equation}
\pi_{i 2 \mid 1}  = n  w_{i1}\,,\;\; i \in B\,,
\label{cond1}
\end{equation}
where $w_{i1}$ is the importance weight in (\ref{5}). To guarantee  
\begin{equation}
\pi_{i 2 \mid 1} \in (0,1] \,,\;\; i \in B\,,
\label{cond2}
\end{equation}
 we should choose $n\le  1/\max_{i \in B}\{ w_{i1}\} $.  Once $\{\pi_{i2 \mid 1}:i\in B\}$ satisfying (\ref{cond1}) and (\ref{cond2}) are found, we can  apply any unequal probability sampling techniques to obtain the second-phase sample;
see Till\'{e} (2006) for details on   algorithms for unequal probability sampling designs.

Once the second-phase sample $B_2$ is obtained, we can use the sample mean of $y_i$ in $B_2$ to estimate $\theta$. The variance estimator of $\hat{\theta}_{B2}$ can be decomposed as
$$ \mbox{Var}( \hat{\theta}_{B2}) = \mbox{Var}( \hat{\theta}_{B1} ) + \mbox{Var} ( \hat{\theta}_{B2} - \hat{\theta}_{B1})\,, $$
where the first term is of order $O(N_B^{-1})$, and the second term is of order $O(n^{-1})$. If $n/N_{B}=o(1)$, the first term can be safely ignored, and we only need to estimate the second term.  Since we can express
$$ \hat{\theta}_{B2} = \sum_{i \in B_2} \frac{1}{ \pi_{i2 \mid 1} } (w_{i1} y_i)\,,$$
we can apply the standard variance estimation formula for the Horvitz--Thompson estimator (Horvitz and Thompson, 1952) by treating the big data as the finite population. That is, we can use
$$ \hat{V}  = \sum_{i \in B_2} \sum_{j \in B_2} \frac{ \pi_{ij 2 \mid 1} - \pi_{i 2 \mid 1} \pi_{j2 \mid 1} }{ \pi_{ij 2 \mid 1} }
\frac{ w_{i1} y_i}{ \pi_{i 2 \mid 1} } \frac{ w_{j1} y_j }{ \pi_{j 2 \mid 1}} $$
as a variance estimator for $\hat{\theta}_{B2}$, where $\pi_{ij 2 \mid 1}$ is the joint inclusion probability for the second-phase sampling.

\section{Data integration}

Survey data integration is an emerging area of research, which aims to combine information from two independent surveys from the same target population.
Kim {\em et al.} (2016) propose a new method of survey data integration using fractional imputation of Kim (2011) under the instrumental variable assumption, and  Park  {\em et al.} (2017) use a measurement error model to combine information from two independent surveys.

Survey data integration idea can be used to combine big data with survey data. Here, we assume that we have two data sources, one is a survey data (denoted by $A$) and the other is a big data (denoted by $B$) which is subject to selection bias. We assume that item $\bx$ is available from survey data while $(\bx,y)$ is available from the big data, and $n/N_B=o(1)$, where $n$ is the sample size of $A$. We are interested in estimating the population mean $\bar{Y}_N$ by combing two data sources. Because of the selection bias, the sample mean  $\bar{Y}_B$ from the big data is biased. Table 1 presents the data structure for this setup.

If both samples were probability samples, then synthetic data imputation can be used to create imputed values of $y_i$ in the sample A. Such synthetic data imputation, or mass imputation, is also considered by Legg and Fuller (2009) and Kim and Rao
(2011).
When $B$ is a non-probability sample,  Rivers (2007) proposes a mass imputation approach using nearest neighbor imputation for survey integration. That is, we can use $\bx$ to find the nearest neighbor in the big data sample $B$ to create an imputed value of $y_i$  for each element in the sample $A$. Once the imputed values of $y_i$ are created for all the elements in the sample $A$, we can compute an imputed estimator of $\theta=E(Y)$ from the sample $A$. Such a method can be justified if
\begin{equation}
f_B( y \mid \bx) = f( y \mid \bx )\,, \label{eq: MAR}
\end{equation}
where $f_B(y\mid \bx)$ is the conditional density of $y$ given $\bx$ for the big data sample $B$, and $f( y \mid \bx )$ is that for the target population.
This assumption, which is called transportability, can be achieved if the selection mechanism for big data is non-informative (Pfeffermann, 1993). Because the sample $A$ is a probability sample, the imputation estimator $\hat{\theta}_{A,I} = N^{-1}\sum_{i\in A}d_iy_i^*$ is approximately  unbiased under certain conditions, where $y_i^*$  is the imputed value of unit $i$, and $d_i$ is the associated sampling weight.

Instead of using mass imputation of Rivers (2007), we propose to use propensity score weighting for the big data based on auxiliary information in the sample $A$. To formally describe the idea, we first assume that we can observe $\delta_i$, the big data sample inclusion indicator, from the sample $A$. That is, among the elements in the sample $A$,  it is possible to obtain the membership information from the big data sample $B$. For example, if  the big data sample $B$ consists of people  using a certain credit card, then we can obtain $\delta_i$ from $A$ by asking  whether person $i$ uses the credit card.

We assume that the selection mechanism of the big data sample is ignorable
\begin{equation*}
P ( \delta_i =1 \mid \bx_i, y_i) =  P( \delta_i =1 \mid \bx_i)\,,\;\; i \in U\,,
\label{11}
\end{equation*}
and it follows a parametric model
\begin{equation}
P ( \delta_i =1 \mid \bx_i ) = p_i( \bflam)  \in (0, 1]\,,\;\; i \in U\,,
\label{12}
\end{equation}
where $p_i( \bflam) = p(\bx_i^\Ti \bflam)$
for some known function $p( \cdot)$ with second continuous derivatives with respect to an unknown parameter $\bflam$, and $p_i(\bflam)^{-1} =O(N)$. Since we observe $(\delta_i, \bx_i)$ from the sample $A$, we can estimate $\bflam$ by maximizing the pseudo log-likelihood function of $\bflam$ given by
\begin{equation*}
l ( \bflam ) = \sum_{i \in A} d_{i} [  \delta_i \log \{p_i( \bflam)\} + (1-\delta_i) \log\{1-p_i(\bflam)\} ] \,.
\label{13}
\end{equation*}
Once the pseudo maximum likelihood estimator $\hat{\bflam}$ is obtained, then we can use a propensity score weighting estimator, that is,
\begin{equation}
\hat{\theta}_{B, PS} = \frac{ \sum_{i \in B} {p}_i(\hat{\bflam})^{-1} y_i }{ \sum_{i \in B} {p}_i(\hat{\bflam})^{-1} }
\label{14}
\end{equation}
as a weighted estimator of $\theta$ from the big data sample $B$.

%  {Because (\ref{14}) uses the big data sample, we expect that $\hat{\theta}_{B, PS}$ in (\ref{14}) is more efficient than the imputation estimator of Rivers (2007). The second simulation study in Section 5 will show the result.  }

To discuss variance estimation of $\hat{\theta}_{B, PS}$, note that $(\hat{\bflam},\hat{\theta}_{B, PS})'$ is a solution to the joint estimating equation, that is,
\begin{eqnarray}
&U( \theta, \bflam) \equiv  \sum_{i \in B} p_i( \bflam)^{-1} (y_i - \theta)  = 0\,,  \label{eq: theta}\\
&S( \bflam)  \equiv  \sum_{i \in A} d_{i} \{ \delta_i -p_i( \bflam) \} g_i( \bflam) = 0\,,   \label{eq: lambda}
\end{eqnarray}
where
$g_i (\bflam) = \partial \mbox{logit}  \{p_i( \bflam)  \}/ \partial \bflam $. Thus, by
using the sandwich formula, we can obtain a consistent variance estimator of $\hat{\theta}_{B,PS}$; see Appendix A for details.

\begin{remark}
	If we can build a working outcome regression model for $E(Y \mid \bx)$, say $E(Y \mid \bx ) = \bx^\Ti \bfbeta$, we can construct a doubly robust estimator (Kim and Haziza, 2014) given by
	\begin{equation}
	\hat{\theta}_{B, DR} = \frac{1}{N} \left\{  \sum_{i \in B} \frac{1}{ p_i ( \hat{\bflam} ) } \left( y_i - \bx_i^\Ti \hat{\bfbeta} \right) + \sum_{i \in A} d_i \bx_i^\Ti \hat{\bfbeta} \right\} \,,\label{eq: doubly robust}
	\end{equation}
	where $\hat{\bfbeta}$ is the estimated regression coefficient based on the big data sample.  We assume that an intercept term is included in $\bx$. Under the model assumption (\ref{eq: MAR}), $\hat{\bfbeta}$ can be obtained by ordinary least squares. To show double robustness, let $\hat{\theta}_{A,HT} = N^{-1}\sum_{i\in A}d_iy_i$ be the Horvitz--Thompson estimator  of $\theta$ from the sample $A$. Note that 
	$$
	\hat{\theta}_{B,DR} - \hat{\theta}_{A,HT} = \frac{1}{N}\left\{ \sum_{i\in B} \frac{1}{ p_i ( \hat{\bflam} ) } \hat{e}_i - \sum_{i\in A}d_i \hat{e}_i\right\}\,,
	$$ where $\hat{e}_i = y_i - \bx_i^\Ti\hat{\bfbeta}$. Thus, if the model (\ref{12}) is correctly specified, we have 
	\begin{eqnarray}
	E_\delta( \hat{\theta}_{B,DR} - \hat{\theta}_{A,HT}) &\approx& \frac{1}{N}\left(\sum_{i\in U}e_i - \sum_{i\in A}d_ie_i\right)\,,\label{15}
	\end{eqnarray} 
	where $e_i = y_i - \bx_i^\Ti\bfbeta^*$ and $\bfbeta^*$ is the probability limit of $\hat{\bfbeta}$. The right side of (\ref{15}) is design-unbiased to zero, so $\hat{\theta}_{B,DR}$ is asymptotically unbiased under model (\ref{12}). On the other hand, if $E(Y\mid \bx) = \bx^\Ti\bfbeta$ is correctly specified, then, 
	\begin{eqnarray*}
		\frac{1}{N}E\left\{\sum_{i\in B}\frac{1}{ p_i ( \hat{\bflam} ) } \hat{e}_i \mid  B\right\} &\approx& \frac{1}{N}\sum_{i\in B}\frac{1}{ p_i ( {\bflam}^* ) } E(\hat{e}_i \mid B)\,, \label{16} \\
		\frac{1}{N}E\left(\sum_{i\in A}d_i \hat{e}_i \mid  B\right) &=& \frac{1}{N}\sum_{i\in U} E\left(\hat{e}_i \mid  B\right) \, ,\label{17}  
	\end{eqnarray*} 
	where $ \hat{e}_i = y_i - \bx_i^\Ti\hat{\bfbeta} $ and $\bflam^*$ is the probability limit of $\hat{\bflam}$. Note that $E(\hat{e}_i\mid B) = 0$ under $E(Y\mid \bx) = \bx^\Ti\bfbeta$ and MAR.  
	Thus, we have 
	\begin{equation}
	E(\hat{\theta}_{B,DR} - \hat{\theta}_{A,HT})\approx 0\,,\label{18}
	\end{equation} if the outcome regression model is correctly specified. Therefore, we have established  double robustness of $\hat{\theta}_{B,DR}$.  Variance estimation of $\hat{\theta}_{B, DR}$ is discussed in Appendix B.

\end{remark}
%[Variance estimation needs to be discussed. ]

\section{Simulation Study}

\subsection{Inverse sampling}
In this simulation study, we consider the proposed inverse sampling under a simple setup.
A finite population is generated by
\begin{eqnarray*}
	&y_i = 5+3x_i + e_i\,,\;\; i =1,\ldots,N\,,
\end{eqnarray*}
 where $x_i \sim  \mbox{Exp}(1)$, $ e_i \sim N(0,x_i^2)$, $N=1,000,000$, $N(\mu,\sigma^2)$ is a normal distribution with mean $\mu$ and variance $\sigma^2$, and $\mbox{Exp}(\lambda)$ is an exponential distribution with mean $\lambda$. The inclusion indicator of the big data sample is generated by
$
\delta_i \sim \mbox{Ber}(p_i)
$ independently
for $i=1,\ldots,N$, where $\mbox{logit}(p_i) =  \phi (x_i-2)$, $\mbox{Ber}(p)$ is a Bernoulli distribution with success probability $p$, and $\mbox{logit}(x) = \log(x) - \log(1-x)$ for $x\in(0,1)$. In addition, we assume that the population mean $\bar{X}_N$ is known. We consider two cases, $\phi = -0.2$ and $\phi=-0.5$, and we are interested in making inference for the population mean $\bar{Y}_N$ and a proportion $P_N = N^{-1}\sum_{i=1}^N I (y_i<6)$, where $I(x<a) = 1$ if $x<a$ for a given number $a$, and 0 otherwise.

We compute the following three estimators with  $n=500$ and $n=1,000$, respectively, and recall that $n$ is the sample size for the second-phase sampling.
\begin{itemize}
	\item[I.] Naive estimator: We use simple random sampling to get a sample of size $n$ from the big data sample $B$.
	\item[II.] Calibration estimator: From the sample obtained by the naive method, we use the exponential tilting method described in Section 3 to obtain a calibration estimator using $\bar{X}_N$ information.
	\item[III.] Proposed inverse sampling estimator: First, we obtain the  important weights   in  (\ref{8}) satisfying the calibration condition (\ref{9}), and then  a sample of size $n$ is selected by probability-proportional-to-size sampling.
	
	%	\item[Method IV.] Proposed-calibration method. Based on the sample obtained by the proposed method, we use the exponential tilting to calibrate the estimator by $\bar{X}_N$.
\end{itemize}

We conduct 10,000 Monte Carlo simulations and compare the three estimators with respect to the bias and standard error of the point estimator, the relative bias of the estimated standard error and the coverage rate of a 95\% confidence interval obtained from the Wald-type method. Table \ref{tab: simulation 1} summarizes the simulation results. The naive estimator works poorly since it does not account for the selection bias of the big data sample. Specifically, its coverage rate decreases as the sample size gets larger, conforming the big data paradox of Meng (2018). Although the calibration estimator works better than the naive one by incorporating external information, its performance  is still questionable since its variance estimator is biased when $\phi=-0.5$, that is, when the mean of the big data sample, $N_B^{-1}\sum_{i\in B}x_i$, differs significantly from $\bar{X}_N$. For estimating $\bar{Y}_N$, which is a linear function of $\bar{X}_N$ in our simulation, the biases of the calibration estimator and the proposed inverse sampling estimator are negligible compared with the standard errors, and the coverage rates of these two methods are close to 0.95 in spite of the small bias of the estimated variance of the   calibration estimator. For estimating $P_N$, which is not a linear function of $\bar{X}_N$, the biases of the  calibration estimator and the proposed inverse sampling estimator are approximately the same, but they are not negligible compared with the standard error when $\phi=-0.5$. Thus, the coverage rates of the calibration estimator and the proposed inverse sampling estimator are below 0.95. Besides, variance estimator of the proposed inverse sampling estimator is unbiased for all cases, but that  of the calibration estimator becomes worse when $\phi=-0.5$.

\subsection{Data integration}
We use a  simulation setup similar to Kim and Haziza (2014) to  compare the two proposed estimators shown in (\ref{14}) and (\ref{eq: doubly robust}) with a naive estimator and Rivers' method.
We consider the following two outcome regression   models for generating the finite population. 
\begin{enumerate}
	\item[I.] Linear model. That is, 
	\begin{equation}
	y_i = 1 + x_{1,i} + x_{2,i} + \epsilon_i\,,\;\; i=1,\ldots,N\,,\label{eq: simu 2 linear model}
	\end{equation}
where $x_{1,i}\sim N(1,1)$, $x_{2,i}\sim \mbox{Exp}(1)$, $\epsilon_i\sim N(0,1)$, $N=1,000,000$, and $(x_{1,i},x_{2,i},\epsilon_{i})$ is pair-wise independent.
	\item[II.] Nonlinear model. That is,
	\begin{equation}
	y_i = 0.5(x_{1,i} - 1.5)^2 + x_{2,i} + \epsilon_i\,,\;\; i=1,\ldots,N\,, \label{eq: simu 2 nonlinear model}
	\end{equation}
	where $(x_{1,i},x_{2,i},\epsilon_i)$ is the same with those in the linear model.
\end{enumerate}
The sampling indicator of the big data sample is generated by 
$
\delta_i \sim \mbox{Ber}(p_i)
$ independently
for $i=1,\ldots,N$,
and we consider the following two big data propensity models.  
\begin{enumerate}
	\item[I.] Linear logistic model. That is,
	\begin{equation}
	\mbox{logit}(p_i) =  x_{2,i}\,,\;\; i=1,\ldots,N\,.\label{eq: simu 2 res lin}
	\end{equation}
	
	\item[II.] Nonlinear logistic model. That is,
	\begin{equation}
	\mbox{logit}(p_i) =  -0.5+0.5(x_{2,i}-2)^2\,,\;\; i=1,\ldots,N\,.\label{eq: simu 2 res non}
	\end{equation}
	
\end{enumerate}
The average sampling rates for the big data are about 60\% under both models.

We consider the following three scenarios to generate the finite population and the big data sample. 
\begin{enumerate}
	\item[I.] Both the outcome regression model and the big data propensity model are linear. That is, the finite population is generated by (\ref{eq: simu 2 linear model}), and the sampling indicator of the big data sample is generated by (\ref{eq: simu 2 res lin}).
	\item[II.] The outcome regression  model is linear, and a nonlinear logistic model is used for the  big data propensity model. That is, we use (\ref{eq: simu 2 linear model}) to generate the finite population, and use (\ref{eq: simu 2 res non}) to generate the sampling indicator of the big data sample.
	\item[III.] The outcome regression model is nonlinear, and the  big data propensity model is linear. That is, we use (\ref{eq: simu 2 nonlinear model}) and (\ref{eq: simu 2 res lin}) to generate the finite population and big data sample. 
\end{enumerate}
The parameter of interest is the population mean $\bar{Y}_N$.
We use simple random sampling to get an independent sample $A$ of size $n$, and we consider $n=500$ and $n=1,000$. We compare the following methods for estimating $\bar{Y}_N$ and the corresponding 95\% confidence interval.
\begin{enumerate}
	\item[I.] Naive estimator. We use sample mean and sample variance of the big data sample to make inference.
	\item[II.] Rivers' method. The nearest neighbor is obtained by the Euclidean norm based on $(x_{1,i},x_{2,i})$.
	\item[III.] The proposed propensity score (PS) weighting estimator (\ref{14}) using a logistic model for  $p(\cdot)$, that is,
	$
	\mbox{logit}\{p_i(\bflam)\} = \lambda_0 + \lambda_1x_{2,i}.
	$
	\item[IV.] The proposed doubly robust (DR) estimator in  (\ref{eq: doubly robust}). The working outcome regression model is $E(y_i\mid x_{1,i},x_{2,i}) = \beta_0 + \beta_1x_{1,i} + \beta_2x_{2,i}$, and the working  big data propensity model is the same as that in Method III.
\end{enumerate}

For each scenario, we conduct 2,000 Monte Carlo simulations to compare the data integration estimators regarding the bias and standard error of the point estimator and the coverage rate of the 95\% confidence interval obtained by the Wald-type method. Table \ref{tab: simulation 2} summarizes the simulation results. The naive estimator is biased since it does not account for the random mechanism for the big data sample, and its coverage rate is zero for all scenarios. Rivers' method works well in terms of the bias and coverage rate in all three scenarios. For Scenario I and Scenario III, the proposed PS estimator has the smallest standard error compared with others, and its bias and coverage rate is as good as those by Rivers' method and the proposed DR estimator. However, the proposed PS estimator is sensitive to the mis-specification of the big data propensity model, and its estimates are biased in Scenario II, where a nonlinear logistic model is used for the big data propensity model.  For all three scenarios, the proposed DR estimator works better than the Rivers' method in terms of the standard errors, and both methods have approximately the same bias and coverage rate. 

\begin{remark}
	The asymptotic variance of the Rivers' method is $\sigma_y^2/n$ (Rivers, 2007), and it is consistent with the simulation results shown in Table \ref{tab: simulation 2} for all scenarios, where  $\sigma_y^2$ is the variance of $y$ with respect to the outcome regression model,  $\sigma_y^2 = 3$ for Scenario I and II and $\sigma_y^2 = 2.75$ for Scenario III.  For the proposed DR estimator, if one of the  working outcome regression model and the working big data propensity model is correctly specified,  the variance of $\hat{\theta}_{B, DR}$ can be  estimated by the sampling variance of the imputed values $\{\bx_i^\Ti{\bfbeta}^*:i\in A\}$, which is $V_{B,DR}={\bfbeta^*_x}^\Ti \Sigma_{xx}{\bfbeta^*_x}/n$, where $\bfbeta_x^*$ is the  coefficient of $(x_{1,i},x_{2,i})$ in $\bfbeta^*$ shown in Remark 1, and $\Sigma_{xx}$ is the variance of $(x_{1,i},x_{2,i})$; see Appendix B for details. For Scenario I and Scenario II, $V_{B,DR} = 2/n$ and $V_{B,DR} \approx 1.25/n$ for Scenario III, and the results are consistent with those shown in Table \ref{tab: simulation 2}. Thus, the proposed DR estimator is more efficient than the Rivers' method in all three scenarios.
\end{remark}

\section{Conclusion}

Adjusting for the selection bias in big data is an important practical problem. By properly incorporating the auxiliary information from an external source, we can reduce the selection bias either by inverse sampling or by propensity score weighting. Doubly robust estimation shows good performance in the simulation study, and extension to multiple robust estimation (Chen and Haziza, 2017) seems to be a promising research area. The proposed methods implicitly assume that the selection mechanism for big data is missing at random (MAR) in the sense of Rubin (1976). If MAR assumption does not hold, then we can build a Not-Missing-At-Random model for the selection mechanism and estimate the model parameters (Chang and Kott, 2008; Riddles {\em et al.}, 2016).

If there is error in the
matching mechanism, then misclassification errors for $\delta$ can arise, and capture-recapture experiments
(Chen and Kim, 2014) can be useful in this situation. Such extensions will be topics for future research.

\section*{Acknowledgment}
The authors wish to thank Professors J.~N.~K. Rao and   Shu Yang for their constructive comments. The research was partially supported by a grant from U.S. National Science Foundation.

\appendix
\renewcommand{\theequation}{A.\arabic{equation}}
\setcounter{equation}{0}
\section*{Appendix}
\subsection*{A. Variance estimation of $\hat{\theta}_{B,PS}$ in (\ref{14}) }
\label{app: variance estimation}
We rewrite (\ref{eq: theta}) and (\ref{eq: lambda}) as 
\begin{eqnarray}
&U( \theta, \bflam) =  \sum_{i=1}^N \delta_ip_i( \bflam)^{-1} (y_i - \theta)\,,  \notag\\
&S( \bflam) = \sum_{i=1}^N I_id_{i} \{ \delta_i -p_i( \bflam) \} g_i( \bflam)\,,   \notag
\end{eqnarray}
where $I_i$ is the sampling indicator for sample $A$, $I_i = 1$ if $i\in A$ and 0 otherwise and $g_i (\bflam) = \partial \mbox{logit}  \{p_i( \bflam)  \}/ \partial \bflam $.
Then, we have 
\begin{eqnarray}
\mbox{Var}\{U( \theta, \bflam)\} &=&\sum_{i=1}^N\{1-p_i( \bflam)\}p_i( \bflam)^{-1} (y_i - \theta)^2\,,\label{eq: var u a}\\
\mbox{Var}\{S( \bflam)\} &=& E\left[\mbox{Var}\{S( \bflam)\mid A\}\right] + \mbox{Var}\left[E\{S( \bflam)\mid A\} \right]\notag \\ 
&=& E\left[\mbox{Var}\{S( \bflam)\mid A\}\right]\,, \label{eq: var s a}\\
\mbox{Cov}\{U( \theta, \bflam),S( \bflam)\} &=& E\left[\mbox{Cov}\{U( \theta, \bflam),S( \bflam)\mid A\}\right]  + \mbox{Cov}\left[ E\{U( \theta, \bflam)\mid A\},E\{S( \bflam)\mid A\} \right]\notag \\ 
&=& E\left[\mbox{Cov}\{U( \theta, \bflam),S( \bflam)\mid A\}\right]\,, \label{eq: cov a}
\end{eqnarray}
where (\ref{eq: var u a}) holds since $\{\delta_i:i\in U\}$ are pair-wise independent, the second equalities of (\ref{eq: var s a}) and (\ref{eq: cov a}) hold since $\delta_i$ is independent with $I_i$, and $\mbox{Cov}\{U( \theta, \bflam),S( \bflam)\mid A\} = \sum_{i=1}^N (y_i-\theta)I_id_i \{1-p_i(\bflam)\}g_i(\bflam)^\Ti$. 

Therefore, we can estimate (\ref{eq: var u a}) to (\ref{eq: cov a}) by 
\begin{eqnarray}
\hat{V}\{U( \theta, \bflam)\}  &=& \sum_{i\in B}\{1-p_i( \bflam)\}p_i( \bflam)^{-2} (y_i - \theta)^2\,,\label{eq: var u approx} \\ 
\hat{V}\{S( \bflam)\} &=& \hat{V}\{S( \bflam)\mid A\} \notag \\
&=& \sum_{i\in A}d_i^2p_i(\bflam)\{1-p_i(\bflam)\}g_i(\bflam)^\Ti g_i(\bflam)\,,
\label{eq: var s approx}\\
\hat{C}\{U( \theta, \bflam),S( \bflam)\} &=& \hat{C}\{U( \theta, \bflam),S( \bflam)\mid A\}\notag \\ 
&=& \sum_{i\in A\cap B} d_ip_i(\bflam)^{-1}(y_i-\theta)\{1-p_i(\bflam)\}g_i(\bflam)^\Ti\,.\label{eq: var c approx}
\end{eqnarray}

Denote 
$$
H({\theta},{\bflam}) = \begin{pmatrix}
\frac{\partial U(\theta,\bflam)}{\partial \theta^\Ti} & 	\frac{\partial U(\theta,\bflam)}{\partial \bflam^\Ti} \\ 
0 & \frac{\partial S(\bflam)}{\partial \bflam^\Ti}
\end{pmatrix}
$$
to be the Hessian matrix of $[U(\theta,\bflam)^\Ti,S(\bflam)^\Ti]^\Ti$, and 
$$
\hat{V}_{U,S}(\hat{\theta},\hat{\bflam}) = \begin{pmatrix}
\hat{V}\{U( {\theta},{\bflam})\} & 	\hat{C}\{U( {\theta},{\bflam}),S( {\bflam})\} \\
\hat{C}\{U( {\theta},{\bflam}),S( {\bflam})\}^\Ti & \hat{V}\{S( {\bflam})\}
\end{pmatrix}
$$
to be the variance estimator of $\{U(\theta,\bflam),S(\bflam)\}$ based on (\ref{eq: var u approx}) to (\ref{eq: var c approx}).

Thus, by the sandwich formula, the variance of $(\hat{\theta}_{B,PS},\hat{\bflam})$ can be estimated by 
\begin{equation}
H(\hat{\theta},\hat{\bflam})^{-1} \hat{V}_{U,S}(\hat{\theta},\hat{\bflam})\{H(\hat{\theta},\hat{\bflam})^{-1}\}^{\Ti}\,,\label{eq: append a last}
\end{equation}
where $\hat{\theta} = \hat{\theta}_{B,PS}$, and the variance estimator of $\hat{\theta}_{B,PS}$ is the (1,1)-th element of (\ref{eq: append a last}).

\renewcommand{\theequation}{B.\arabic{equation}}
\setcounter{equation}{0}
\subsection*{B. Variance estimation of the double robust estimator}\label{app: var dou}
Denote
\begin{eqnarray*}
	\tilde{\theta}_{B, DR}(\hat{\bflam}) =  \frac{1}{N} \left\{  \sum_{i \in B} \frac{1}{ p_i ( \hat{\bflam} ) } \left( y_i - \bx_i^\Ti {\bfbeta}^* \right) + \sum_{i \in A} d_i \bx_i^\Ti {\bfbeta}^* \right\}\,,
\end{eqnarray*}
where $\bfbeta^*$ is the probability limit of $\hat{\bfbeta}$. Since $\mbox{Var}(\hat{\bfbeta}) = O(N_B^{-1})$, $\tilde{\theta}_{B, DR}(\hat{\bflam})$ is asymptotically equivalent to $\hat{\theta}_{B,DR}(\hat{\bflam})$ if $n/N_B=o(1)$.

Let $\bflam^*$ be the probability limit of $\hat{\bflam}$, and we have
\begin{eqnarray}
\tilde{\theta}_{B, DR}(\hat{\bflam}) =N^{-1}\sum_{i\in B}\frac{1}{p_i(\bflam^*)}(y_i-\bx_i^\Ti\bfbeta^*)+ \eta_B(\bflam^*)^\Ti(\hat{\bflam} - \bflam^*) + \hat{\theta}_{A,reg} + o_p(n^{-1/2})\label{eq: append b}
\end{eqnarray}
by Taylor expansion,
where $\eta_B(\bflam^*) = N^{-1}\sum_{i\in B}p_i(\bflam^*)^{-1}\{p_i(\bflam^*)-1\}(y_i - \bx_i^\Ti{\bfbeta}^*)\bx_i$ and $\hat{\theta}_{A,reg} = N^{-1}\sum_{i\in A}d_i \bx_i^\Ti {\bfbeta}^*$. 

Note that $\hat{\bfbeta}$ is a consistent estimator of $\bfbeta^*$. Under the model assumption (\ref{eq: MAR}), $\bfbeta^*$ is also the probability limit of $\bfbeta_N$, where 
$\bfbeta_N$ solves $\sum_{i=1}^N(y_i-\bx_i^\Ti\bfbeta)\bx_i=0$. Thus, we have
\begin{eqnarray}
\hat{\bfbeta}&=&\bfbeta^* + O_p(N_B^{-1/2})\,,\label{eq: inequality 1}\\
\bfbeta_N&=&\bfbeta^* + O_p(N_B^{-1/2})\,,\label{eq: inequality 2}
\end{eqnarray}
where the second result holds since ${\bfbeta}_N=\bfbeta^* + O_p(N^{-1/2})$ and $N_B/N=O(1)$.
Next, we wish to show 
\begin{equation}
\eta_B(\bflam^*) = O_p(N_B^{-1/2})\,, \label{eq: append B eta B}
\end{equation}
if one of the outcome regression model and the big data propensity model is correctly specified. 
Suppose that the outcome regression model is correctly specified. Then, ${e}_i = y_i - \bx_i^\Ti{\bfbeta}^*$ is independent with $\bx_i$, so (\ref{eq: append B eta B}) holds under mild conditions on the working big data propensity model.

If the big data propensity model is correctly specified, consider $$\eta_B(\bflam^*)=N^{-1}\sum_{i\in B}(y_i-\bx_i^\Ti\bfbeta^*)  - N^{-1}\sum_{i\in B}p_i(\bflam^*)^{-1}(y_i-\bx_i^\Ti\bfbeta^*)=\eta_{B,1}(\bflam^*) - \eta_{B,2}(\bflam^*)\,. $$

First, note that $\sum_{i\in B}(y_i-\bx_i^\Ti\hat{\bfbeta})=0$, and we have
\begin{eqnarray}
\eta_{B,1}(\bflam^*)=N^{-1}\sum_{i\in B}(y_i-\bx_i^\Ti\bfbeta^*) &=& 	N^{-1}\sum_{i\in B}(y_i-\bx_i^\Ti\hat{\bfbeta}) + N^{-1}\sum_{i\in B}\bx_i^\Ti(\hat{\bfbeta}-{\bfbeta}^*)\notag \\ &\leq& O_p(N_B^{-1/2})N^{-1}\sum_{i=1}^N \lVert \bx_i \rVert_2\notag \\ 
&=& O_p(N_B^{-1/2})\,, \label{eq: append B3}
\end{eqnarray}
where the inequality holds by (\ref{eq: inequality 1}), and the second equality holds if $\bx_i$ has a finite second moment. Now, to discuss $\eta_{B,2}(\bflam^*)$, note that $\bfbeta_N$ satisfies 
$\sum_{i=1}^N(y_i - \bx_i^\Ti\bfbeta_N)=0$. Thus,
\begin{eqnarray}
\eta_{B,2}(\bflam^*)&=& N^{-1}\sum_{i\in B}p_i(\bflam^*)^{-1}(y_i-\bx_i^\Ti\bfbeta^*)\notag \\ 
&=& N^{-1}\sum_{i=1}^N(y_i-\bx_i^\Ti\bfbeta^*)+O_p(N_B^{-1/2})\notag \\ 
&=&N^{-1}\sum_{i=1}^N (y_i - \bx_i^\Ti{\bfbeta}_N) + N^{-1}\sum_{i=1}^N\bx_i^\Ti (\bfbeta_N-{\bfbeta}^*)+O_p(N_B^{-1/2}) \notag \\ 
&=& O_p(N_B^{-1/2})\,, \label{eq: append B4}
\end{eqnarray}
where the last equality holds by (\ref{eq: inequality 2}).
Thus, if the big data propensity model is correctly specified, we have shown (\ref{eq: append B eta B}) by  (\ref{eq: append B3}) and (\ref{eq: append B4}).

Similarly, we can show that the first term of (\ref{eq: append b}) has order $O_p(N_B^{-1/2})$ if   one of the outcome regression model and the big data propensity model is correctly specified. Thus, the variance of $\tilde{\theta}_{B, DR}(\hat{\bflam})$ can be estimated by 
the sampling variance of $\hat{\theta}_{A,reg}$ under the assumption $n/N_B = o(1)$.

\section*{References}

\bigskip

\def\beginref{\begingroup
	\clubpenalty=10000
	\widowpenalty=10000
	\normalbaselines\parindent 0pt
	\parskip.0\baselineskip
	\everypar{\hangindent1em}}
\def\endref{\par\endgroup}

\newcommand{\harvardand}{\&}
\newcommand{\harvardyearleft}{(}
\newcommand{\harvardyearright}{)}
\beginref

Baker, R., Brick, J.M., Bates, N.A., Battaglia, M., Couper, M.P., Dever,
J.A., Gile, K.J. \harvardand\ Tourangeau, R.  \harvardyearleft
2013\harvardyearright.
Summary report of the {AAPOR} task force on non-probability sampling,
{\em J. Surv. Stat. Methodol.}, {\bf 1}, 90--143.

Bethlehem, J.  \harvardyearleft 2016\harvardyearright.
Solving the nonresponse problem with sample matching?, 
{\em Soc. Sci. Comput. Rev.}, {\bf 34}, 59--77.

Breidt, F.J. \harvardand\ Fuller, W.A.  \harvardyearleft
1993\harvardyearright.
Regression weighting for multipurpose sampling, {\em Sankhya B}, {\bf 55}, 297--309.

Chang, T. \harvardand\ Kott, P.S.  \harvardyearleft 2008\harvardyearright.
Using calibration weighting to adjust for nonresponse under a
plausible model, {\em Biometrika}, {\bf 95}, 555--571.

Chen, S. \harvardand\ Haziza, D.  \harvardyearleft 2017\harvardyearright.
Multiply robust imputation procedures for the treatment of item
nonresponse in surveys, {\em Biometrika}, {\bf 104}, 439--453.

Chen, S. \harvardand\ Kim, J.K.  \harvardyearleft 2014\harvardyearright.
Two-phase sampling experiment for propensity score estimation in
self-selected samples, {\em Ann. Appl. Stat.}, {\bf
	8}, 1492--1515.

Cochran, W.G.  \harvardyearleft 1977\harvardyearright.
{\em Sampling Techniques}, 3rd edn, John Wiley \& Sons, New York.

Efron, B. \harvardand\ Hastie, T.  \harvardyearleft 2016\harvardyearright.
{\em Computer Age Statistical Inference}, Cambridge, New York.

Elliott, M. \harvardand\ Valliant, R.  \harvardyearleft 2017\harvardyearright.
Inference for non-probability samples, {\em Stat. Sci.}, {\bf
	32}, 249--264.

Franke, B., Plante, J.-F., Roscher, R., Lee, E.-S.A., Smyth, C., Hatefi, A.,
Chen, F., Gil, E., Schwing, A., Selvitella, A., Hoffman, M.M., Grosse, R.,
Hendricks, D. \harvardand\ Reid, N.  \harvardyearleft 2016\harvardyearright.
Statistical inference, learning and models in big data, {\em
	Int. Stat. Rev.}, {\bf 84}, 371--389.

Fuller, W.A.  \harvardyearleft 2009\harvardyearright.
{\em Sampling Statistics}, John Wiley {\&} Sons, Hoboken.

Goffinet, B. \harvardand\ Wallach, D.  \harvardyearleft 1996\harvardyearright.
Optimized importance sampling quantile estimation, {\em Biometrika}
, {\bf 83}, 791--800.

Henmi, M., Yoshida, R. \harvardand\ Eguchi, S.  \harvardyearleft
2007\harvardyearright.
Importance sampling via the estimated sampler, {\em Biometrika}, {\bf
	94}, 985--991.

Hidiroglou, M.  \harvardyearleft 2001\harvardyearright.
Double sampling, {\em Surv. Methodol.}, {\bf 27}, 143--154.

Hinkins, S., Oh, H.L. \harvardand\ Scheuren, F.  \harvardyearleft
1997\harvardyearright.
Inverse sampling design algorithms, {\em Surv. Methodol.}, {\bf
	23}, 11--21.

Horvitz, D.G. \harvardand\ Thompson, D.J.  \harvardyearleft
1952\harvardyearright.
A generalization of sampling without replacement from a finite
universe, {\em J. Amer. Statist. Assoc.}, {\bf
	47}(260):~663--685.

Keiding, N. \harvardand\ Louis, T.A.  \harvardyearleft 2016\harvardyearright.
Perils and potentials of self-selected entry to epidemiological
studies and surveys (with discussions), {\em J. Roy. Statist. Soc. Ser. A}, {\bf 179}, 1--28.

Kim, J.K.  \harvardyearleft 2010\harvardyearright.
Calibration estimation using exponential tilting in sample surveys,
{\em Surv. Methodol.}, {\bf 36}, 145--155.

Kim, J.K.  \harvardyearleft 2011\harvardyearright.
Parametric fractional imputation for missing data analysis, {\em
	Biometrika}, {\bf 98}, 119--132.

Kim, J.K., Berg, E. \harvardand\ Park, T.  \harvardyearleft
2016\harvardyearright.
Statistical matching using fractional imputation, {\em Surv. Methodol.}, {\bf 42}, 19--40.

Kim, J.K. \harvardand\ Haziza, D.  \harvardyearleft 2014\harvardyearright.
Doubly robust inference with missing data in survey sampling, {\em
	Statist. Sinica}, {\bf 24}, 375--94.

Kim, J.K., Navarro, A. \harvardand\ Fuller, W.A.  \harvardyearleft
2006\harvardyearright.
Replicate variance estimation after multi-phase stratified sampling,
{\em J. Amer. Statist. Assoc.}, {\bf 101}, 312--320.

Kim, J.K. \harvardand\ Rao, J.N.K.  \harvardyearleft 2011\harvardyearright.
Combining data from two independent surveys: a model-assisted
approach, {\em Biometrika}, {\bf 99}, 85--100.

Legg, J.C. \harvardand\ Fuller, W.A.  \harvardyearleft 2009\harvardyearright
.
Two-phase sampling, {\em in} D. Pfeffermann \harvardand\ C.R. Rao
(eds), {\em Handbook of Statistics 29A, Sample Surveys: Design, Methods and
	Applications}, North Holland, pp. 55--70.

Meng, X.L.  \harvardyearleft 2018\harvardyearright.
Statistical paradises and paradoxes in big data (i): Law of large
populations, big data paradox, and 2016 {US} presidential election.
Submitted.

Park, S., Kim, J.K. \harvardand\ Stukel, D.  \harvardyearleft
2017\harvardyearright.
A measurement error model for survey data integration: combining
information from two surveys, {\em Metron}, {\bf 75}, 345--357.

Pfeffermann, D.  \harvardyearleft 1993\harvardyearright.
The role of sampling weights when modeling survey data, {\em
	Int. Stat. Rev.}, {\bf 61}, 317--337.

Rao, J.N.K. \harvardand\ Molina, I.  \harvardyearleft 2015\harvardyearright.
{\em Small Area Estimation}, 2nd edn, John Wiley \& Sons, Hoboken.

Rao, J.N.K., Scott, A.J. \harvardand\ Benhin, E.  \harvardyearleft
2003\harvardyearright.
Undoing complex survey data structures: some theory and applications
of inverse sampling, {\em Surv. Methodol.}, {\bf 29}, 107--128.

Rao, J.N.K. \harvardand\ Sitter, R.R.  \harvardyearleft
1995\harvardyearright.
Variance estimation under two-phase sampling with application to
imputation for missing data, {\em Biometrika}, {\bf 82}, 453--460.

Riddles, M.K., Kim, J.K. \harvardand\ Im, J.  \harvardyearleft
2016\harvardyearright.
A propensity-score-adjustment method for nonignorable nonresponse,
{\em J. Surv. Stat. Methodol.}, {\bf 4}, 215--245.

Rivers, D.  \harvardyearleft 2007\harvardyearright.
Sampling for web surveys, {\em  Proceedings of the Survey Research Methods Section}, American Statistical Association.

Rubin, D.B.  \harvardyearleft 1976\harvardyearright.
Inference and missing data, {\em Biometrika}, {\bf 63}(3):~581--592.

S\"arndal, C.E., Cassel, C.M. \harvardand\ Wretman, J.H.  \harvardyearleft
1992\harvardyearright.
{\em Model Assisted Survey Sampling}, Springer-Verlag, New York.

Stukel, D. \harvardand\ Kott, P.  \harvardyearleft 1996\harvardyearright.
Jackknife variance estimation under two-phase sampling: An empirical
investigation., {\em Proceedings of the Survey Research Methods Section,
	American Statistical Association}.

Tam, S.-M.  \harvardyearleft 2015\harvardyearright.
A statistical framework for analysing big data, {\em Surv. Statist.}, {\bf 72}, 36--51.

Tam, S.-M. \harvardand\ Clarke, F.  \harvardyearleft 2015\harvardyearright.
Big data, official statistics and some initiatives by the australian
bureau of statistics, {\em Int. Stat. Rev.}, {\bf
	83}, 436--448.

Tam, S.-M. \harvardand\ Kim, J.K.  \harvardyearleft 2018\harvardyearright.
Big data, selection bias and ethics -- an official statistician's
perspective, {\em Stat. J. IAOS}.
Accepted for publication.

Till{\'e}, Y.  \harvardyearleft 2006\harvardyearright.
{\em Sampling Algorithms}, Springer-Verlag, New York.

Till{\'e}, Y.  \harvardyearleft 2016\harvardyearright.
Unequal probability inverse sampling, {\em Surv. Methodol.}, {\bf
	42}, 283--295.

Wu, C. \harvardand\ Sitter, R.R.  \harvardyearleft 2001\harvardyearright.
A model-calibration approach to using complete auxiliary information
from survey data, {\em J. Amer. Statist. Assoc.}, {\bf
	96}, 185--193.
\endref

\newpage
\begin{table}[!t]
	\caption{\label{table2} Data Structure}
	\begin{center}
		\begin{tabular}{rccc}
			 \hline \hline%
			Data   & Representativeness & $X$ & $Y$   \\
			\cline{1-4}
			A & Yes & $\checkmark$  &      \\
			B & No & $\checkmark$ &  $\checkmark$    \\
			%  Sample C & o & \textcolor{red}{o}  &  & o \\
		 \hline \hline
		\end{tabular}
	\end{center}
\end{table}

\newpage

\begin{table}[t]
	\centering
	\caption{Monte Carlo bias (Bias), standard error (SE), relative bias of the estimated standard error (RB.SE) and coverage rate (CR) for different estimators based on 2,000 simulation studies. ``Naive'' stands for the naive estimator,  ``Calibration'' for the calibration estimator, and `` Proposed'' for the proposed inverse sampling estimator. ``Par.'' is short for the parameter that we are interested in.} \label{tab: simulation 1}
	\begin{tabular}{cccrrrrrrrrr}
		 \hline \hline
		\multirow{2}{*}{Par.}	& \multirow{2}{*}{$\phi$} & \multirow{2}{*}{Method}& \multicolumn{4}{c}{$n=500$} &  & \multicolumn{4}{c}{$n=1000$}\\
		&  &                        & Bias & SE& RB.SE& CR   & & Bias & SE& RB.SE& CR\\
		\hline
		\multirow{7}{*}{$\bar{Y}_N$} & \multirow{3}{*}{$-0.2$} & Naive & -0.27 & 0.133 &  0.01 & 0.48 &  & -0.27 & 0.096 & -0.01 & 0.22 \\ 
		&  & Calibration &  0.00 & 0.070 & -0.01 & 0.95 &  &  0.00 & 0.050 & -0.02 & 0.95 \\ 
		&  & Proposed &  0.00 & 0.146 &  0.01 & 0.95 &  &  0.00 & 0.104 &  0.00 & 0.95 \\ 
		&  &  &  &  &  &  &  &  &  &  &  \\ 
		& \multirow{3}{*}{$-0.5$} & Naive & -0.55 & 0.117 &  0.01 & 0.01 &  & -0.55 & 0.083 &  0.00 & 0.00 \\ 
		&  & Calibration &  0.00 & 0.081 & -0.04 & 0.94 &  &  0.00 & 0.057 & -0.03 & 0.94 \\ 
		&  & Proposed &  0.00 & 0.141 &  0.00 & 0.95 &  &  0.00 & 0.101 &  0.00 & 0.95 \\ 
		&  &  &  &  &  &  &  &  &  &  &  \\ 
		\multirow{7}{*}{$P_N$} & \multirow{3}{*}{$-0.2$} & Naive & 0.02 & 0.021 & -0.01 & 0.83 &  & 0.02 & 0.015 & -0.01 & 0.70 \\ 
		&  & Calibration & 0.00 & 0.018 &  0.01 & 0.95 &  & 0.00 & 0.012 &  0.01 & 0.95 \\ 
		&  & Proposed & 0.00 & 0.021 & -0.01 & 0.95 &  & 0.00 & 0.015 &  0.00 & 0.95 \\ 
		&  &  &  &  &  &  &  &  &  &  &  \\ 
		& \multirow{3}{*}{$-0.5$} & Naive &  0.05 & 0.021 & 0.00 & 0.45 &  &  0.04 & 0.015 & 0.01 & 0.16 \\ 
		&  & Calibration & -0.01 & 0.018 & 0.09 & 0.92 &  & -0.01 & 0.012 & 0.09 & 0.89 \\ 
		&  & Proposed & -0.01 & 0.021 & 0.00 & 0.92 &  & -0.01 & 0.014 & 0.02 & 0.90 \\ 
	 \hline \hline
	\end{tabular}
\end{table}

\newpage

\begin{table}[!t]
	\centering
	\caption{Monte Carlo bias (Bias), standard error (SE) and coverage rate (CR) of different data integration methods based on 2,000 simulation studies for each scenario. ``Naive'' stands for the naive estimator, ``Rivers'' for the Rivers' method,  ``PS' for the proposed propensity score weighting estimator and ``DR'' for the proposed doubly robust estimator.} \label{tab: simulation 2}
	
	\begin{tabular}{ccrrrrrrr}
		 \hline \hline
		\multirow{2}{*}{Scenario}&	\multirow{2}{*}{Method}& \multicolumn{3}{c}{$n=500$} &  & \multicolumn{3}{c}{$n=1000$}\\
		& &Bias&SE & CR & &  Bias &SE & CR\\
		\hline
		%\multirow{4}{*}{Scenario I} & Naive & 0.19 & 0.00 & 0.00 &  & 0.19 & 0.00 & 0.00 \\ 
		%& Rivers & 0.00 & 0.08 & 0.95 &  & 0.00 & 0.05 & 0.95 \\ 
		%& PS & 0.00 & 0.02 & 0.95 &  & 0.00 & 0.02 & 0.95 \\ 
		%& DR & 0.00 & 0.06 & 0.95 &  & 0.00 & 0.04 & 0.95 \\ 
		%&  &  &  &  &  &  &  &  \\ 
		%\multirow{4}{*}{Scenario II} & Naive & -0.10 & 0.00 & 0.00 &  & -0.10 & 0.00 & 0.00 \\ 
		%& Rivers &  0.00 & 0.08 & 0.96 &  &  0.00 & 0.06 & 0.94 \\ 
		%& PS &  0.11 & 0.18 & 0.99 &  &  0.08 & 0.09 & 1.00 \\ 
		%&DR &  0.00 & 0.06 & 0.95 &  &  0.00 & 0.05 & 0.95 \\ 
		%&  &  &  &  &  &  &  &  \\ 
		%\multirow{4}{*}{Scenario III} & Naive & 0.19 & 0.00 & 0.00 &  & 0.19 & 0.00 & 0.00 \\ 
		%& Rivers & 0.00 & 0.07 & 0.94 &  & 0.00 & 0.05 & 0.95 \\ 
		%&PS & 0.00 & 0.02 & 0.95 &  & 0.00 & 0.02 & 0.95 \\ 
		%& DR & 0.00 & 0.05 & 0.95 &  & 0.00 & 0.04 & 0.95 \\ 
		\multirow{4}{*}{I} & Naive & 0.19 & 0.001 & 0.00 &  & 0.19 & 0.001 & 0.00 \\ 
		& Rivers & 0.00 & 0.077 & 0.95 &  & 0.00 & 0.054 & 0.95 \\ 
		& PS & 0.00 & 0.023 & 0.95 &  & 0.00 & 0.016 & 0.95 \\ 
		& DR & 0.00 & 0.063 & 0.95 &  & 0.00 & 0.044 & 0.95 \\ 
		&  &  &  &  &  &  &  &  \\ 
		\multirow{4}{*}{II} & Naive & -0.10 & 0.001 & 0.00 &  & -0.10 & 0.001 & 0.00 \\ 
		& Rivers &  0.00 & 0.077 & 0.96 &  &  0.00 & 0.055 & 0.94 \\ 
		& PS &  0.11 & 0.183 & 0.99 &  &  0.08 & 0.085 & 1.00 \\ 
		& DR &  0.00 & 0.063 & 0.95 &  &  0.00 & 0.046 & 0.95 \\ 
		&  &  &  &  &  &  &  &  \\ 
		\multirow{4}{*}{III} & Naive & 0.19 & 0.001 & 0.00 &  & 0.19 & 0.001 & 0.00 \\ 
		& Rivers & 0.00 & 0.074 & 0.94 &  & 0.00 & 0.053 & 0.95 \\ 
		& PS & 0.00 & 0.022 & 0.95 &  & 0.00 & 0.016 & 0.95 \\ 
		& DR & 0.00 & 0.050 & 0.95 &  & 0.00 & 0.035 & 0.95 \\ 
		 \hline \hline
	\end{tabular}
\end{table}

\end{document}